\documentclass[aps,prl,twocolumn,superscriptaddress,groupedaddress,showpacs]{revtex4}
\usepackage{graphicx}
\usepackage{wasysym}
\usepackage[dvips]{color}

\begin{document}
\title{Dual Fermion Method for  Disordered Electronic Systems}



\author{H. Terletska}
\email{terletska.hanna@gmail.com}
\affiliation{Condensed Matter Physics and Materials Science Department,
Brookhaven National Laboratory, Upton, New York 11973, USA }
\affiliation{Department of Physics and Astronomy, Louisiana State University, Baton Rouge, Louisiana 70803, USA}
\author{S.-X. Yang}
\affiliation{Department of Physics and Astronomy, Louisiana State University, Baton Rouge, Louisiana 70803, USA}
\affiliation{Center for Computation and Technology, Louisiana State University, Baton Rouge, Louisiana 70803, USA}
\author{Z. Y. Meng}
\affiliation{Department of Physics and Astronomy, Louisiana State University, Baton Rouge, Louisiana 70803, USA}
\affiliation{Center for Computation and Technology, Louisiana State University, Baton Rouge, Louisiana 70803, USA}
\author{J.\ Moreno}
\affiliation{Department of Physics and Astronomy, Louisiana State University, Baton Rouge, Louisiana 70803, USA}
\affiliation{Center for Computation and Technology, Louisiana State University, Baton Rouge, Louisiana 70803, USA}
\author{M.\ Jarrell}
\affiliation{Department of Physics and Astronomy, Louisiana State University, Baton Rouge, Louisiana 70803, USA}
\affiliation{Center for Computation and Technology, Louisiana State University, Baton Rouge, Louisiana 70803, USA}
\date{\today}

\relpenalty=10000      
\binoppenalty=10000

\begin{abstract}
While the coherent potential approximation (CPA) is the prevalent method for the study of disordered 
electronic systems, it fails to capture non-local correlations and Anderson localization. 
To incorporate such effects, we extend the dual fermion approach to disordered non-interacting 
systems using the replica method. Results for single- and two- particle quantities  show good agreement with 
cluster extensions of the CPA; moreover, weak localization is captured. As a natural extension of the CPA, our 
method presents an alternative to the existing cluster theories. It can be used in various applications, 
including the study of disordered interacting systems, or for the description of non-local effects in 
electronic structure calculations.

\end{abstract}

\maketitle

\paragraph*{Introduction.-}
Disorder, due to doping, impurities or structural defects, is
a common feature of many materials, and may play a crucial role in
determining transport properties. The coherent 
potential approximation (CPA) is the most common theoretical
method used to study disordered systems~\cite{p_soven_67}. While the CPA is 
a successful single-site theory of disorder
it misses, by construction, non-local effects including Anderson localization \cite{Anderson}. 
Cluster extensions of the CPA, such as the Molecular CPA~\cite{tsukada69} and the Dynamical 
Cluster Approximation (DCA) \cite{m_jarrell_01a}, capture non-local correlations within 
the cluster~\cite{a_gonis_92}; however, Anderson localization is still missing.  

The Dual Fermion (DF) formalism~\cite{Rubtsov08},
originally developed  for clean interacting systems, is complementary
to these cluster approaches.  It treats local correlations explicitly in
the ``impurity'' solver, and non-local correlations perturbatively.  So,
if a geometric series of relevant diagrams are included, it has the potential
to capture localization.  Here we present such a DF method for disordered systems.  

The DF formalism is based on a set of auxiliary variables (dual fermions) which are introduced 
into the path integral representation of the lattice partition function via a canonical 
transformation~\cite{{Sarker}, {Tremblay}}.  It maps the lattice onto an impurity embedded in 
a self-consistently determined dual fermion lattice.  The DF lattice problem is treated via 
a perturbation theory involving the DF bare Green function, which is the difference of the 
lattice and impurity Green functions, and the impurity full vertex as the effective bare DF 
interaction~\cite{Janis_parquet}. These features are elegantly incorporated in Rubtsov's 
DF formalism~\cite{Rubtsov08}.

For systems with disorder, the DF mapping has to be done differently. In particular, as 
observable quantities are calculated by derivatives of the free energy, the DF formalism for 
the disorder case then needs to be constructed from the disorder averaged $\left<\ln Z\right>_{av}$ 
instead of from the partition function $Z$ as in~\cite{Rubtsov08}.

In this Letter we employ the replica method \cite{Edwards_Anderson} to deal with such averaging.  
We extend the DF method to systems with disorder, and construct the DF mapping directly on 
the Green function. We demonstrate that our method shows remarkable agreement for the 
single-particle Green function with the results obtained from large cluster DCA calculations. 
Moreover, our disorder DF can account for weak localization in the conductivity. This scheme 
presents a powerful alternative to the existing cluster extensions of CPA, with a broad venue 
of applications, including the possibility of treating both electron-electron interactions and 
disorder on equal footing, or replacing the CPA in electronic structure calculations. 


\paragraph*{Method.-}
The simplest model of disordered electrons is the Anderson model. Its Hamiltonian is
\begin{equation}
H=-\sum_{<ij>}t_{ij}(c_{i}^{\dagger}c_{j}+h.c.)+\sum_{i}V_{i}n_{i},
\end{equation}
where $4 t=1$ sets the energy unit, and the local potential $V_i$  is a site-independent random quantity, 
with a uniform box disorder distribution, $p(V)=\displaystyle \frac{1}{W}\Theta(\frac{W}{2}-|V|)$.


We focus on  the disorder averaged Green function 
\begin{equation}
\left<G_k(w)\right>_{av}=-\frac{\delta}{\delta \eta_{wk}}\left< \ln Z(V_i,\eta_{wk})\right>_{av}|_{\eta_{wk}=0},
\label{GF-defintion}
\end{equation}
with  $\left<(...)\right>_{av}=\int dV p(V) (...)$ standing for the disorder averaging, and $\eta_{wk}$ is a source field. 

In the replica method, one employs the relation $\ln Z=\lim _{m\rightarrow 0} \displaystyle \frac {Z^m-1}{m}$, with $m$ 
being the number of replicas. Hence, the above Green function can be written in terms of $m$ powers of $Z$ 
which is much more tractable than a logarithm. 
%
%
Using Grassmann functional integrals for quantum averaging, and the replica method for disorder averaging, 
one can rewrite Eq.~$(\ref{GF-defintion})$ as ($\mathcal{D}c\equiv\prod_{wk\alpha}{dc^\alpha_{wk}}$, $\alpha$ the replica index)
\begin{equation}
\left<G_{k}(w)\right>_{av}  =  -\lim_{m\rightarrow0}\frac{1}{m}\frac{\delta}{\delta\eta_{wk}} 
\left< \int \mathcal{D}\bar{c} \mathcal{D}c e^{-S[c^{\alpha},\bar{c}^{\alpha}]}\right>_{av}|_{_{\eta_{wk}=0}},
 \label{eq:GF}
 \end{equation}
with the lattice action
\begin{equation}
S  = \sum_{wk\alpha}\bar{c}_{wk}^{\alpha}(-iw_{n}+\varepsilon_{k}-\mu+\eta_{wk})c_{wk}^{\alpha}
+  \sum_{i\alpha}V_{i}\int_{0}^{\beta}d\tau n_{i}^{\alpha}(\tau),
\label{eq:action_original}
\end{equation}
where $w_{n}=(2n+1)\pi T$. Averaging over the distribution $p(V)$ in Eq. $(\ref {eq:action_original})$, we obtain
\begin{equation}
S=\sum_{wk\alpha}\bar{c}_{wk}^{\alpha}(-iw_{n}+\epsilon_k-\mu+\eta_{wk})c_{wk}^{\alpha}+\sum_{i}W(\tilde{n}_{i}),
\label{eq:action_av}
\end{equation}
where $ W(\tilde{n}_{i})$ is the elastic effective interaction between electrons of different replicas, and 
may be expressed through cumulants $<V^{l}>_{c}$ as \cite{m_jarrell_01a}
\begin{eqnarray}
e^{-W(\tilde{n}_{i})} & = & \int dV_{i}p(V_{i})e^{-V_{i}\sum_{\alpha}\int d\tau n_{i}^{\alpha}(\tau)} \nonumber \\
 & = & e^{-\sum_{l=2}^{\infty}\frac{1}{l!}<V^{l}>_{c}\left(\sum_{\alpha}\int d\tau n_{i}^{\alpha}(\tau)\right)^{l}}.
\label{eq:cumulant} 
\end{eqnarray}
%
%
Following the DF procedure of~\cite{Rubtsov08}, we introduce an effective single-site impurity reference problem by 
rewriting the original action as
\begin{equation}
S=\sum _i S_{imp}[c^{\alpha},\bar{c}^{\alpha}] -\sum _{wk\alpha}{\bar{c}^{\alpha}_{wk}(\Delta_{w}-\varepsilon _{k}-\eta_{wk})c^{\alpha}_{wk}},
\label{eq:action_with_imp}
\end{equation}
with an effective impurity action
\begin{equation}
S_{imp}=\sum_{\alpha w}\bar{c}^{\alpha}_{i w}(-iw-\mu+\Delta_{w})c^{\alpha}_{i w}+W(\tilde{n}_{i}),
\end{equation}
where $\Delta_w$ is a local, and yet unknown, hybridization function describing the interaction of the impurity with 
the effective medium. Our goal is to express the Green function and other properties of the original 
lattice via the quantities of such impurity problem.

So far we have moved the local part of the lattice action to the effective impurity. One can go further and 
transfer the  non-local part of the action to the auxiliary degrees of freedom, so that the original real 
fermions carry information about the local part only. In the DF scheme~\cite{Rubtsov08} this is done by 
introducing the auxiliary dual fermions via Gaussian transformation of the non-local part of 
Eq. $(\ref {eq:action_with_imp})$, i.e.,
%
%
\begin{equation}
e^{\bar{c}_{wk}^{\alpha}A_{wk}^{2}c_{wk}^{\alpha}}  = \frac{A_{wk}^{2}}{\lambda_{w}^{2}}
\int \mathcal{D}\bar{f}\mathcal{D}f e^{-\lambda_{w}(\bar{c}_{wk}^{\alpha}f_{wk}^{\alpha}+\bar{f}_{wk}^{\alpha}c_{wk}^{\alpha})
-\frac{\lambda_{w}^{2}}{A_{wk}^{2}}\bar{f}_{wk}^{\alpha}f_{wk}^{\alpha}}
\label{eq:Hub-Strat} 
\end{equation}
with $A_{wk}^{2}=(\Delta_{w}-\varepsilon_{k}-\eta_{wk}) $, and $ \lambda_{w} $ yet to be specified.

With such a transformation, the lattice Green function of Eq.~$(\ref {eq:GF})$ can be rewritten as
\begin{eqnarray}
\left<G_{k}(w)\right>_{av} & = & -\lim_{m\rightarrow0}\frac{1}{m}\frac{\delta}{\delta\eta_{wk}}\frac{\left(\Delta_{w}-\varepsilon_{k}-\eta_{wk}\right)}{\lambda_{w}^{2}}\nonumber \\
 & \times & \int \mathcal{D}\bar{f} \mathcal{D}f\,e^{-\sum _{wk\alpha}\lambda_{w}^{2}\bar{f}_{wk}^{\alpha}\left(\Delta_{w}-\varepsilon_{k}-\eta_{wk}\right)^{-1}f_{wk}^{\alpha}}
\nonumber \\
 & \times & \int \mathcal{D}\bar{c} \mathcal{D}c\,e^{-\sum_{i}S_{site}^i[\bar{c}_{i}^{\alpha},c_{i}^{\alpha};\bar{f}_{i}^{\alpha},f_{i}^{\alpha}]}|_{_{\eta_{wk}=0}},
 \nonumber
 \\
\label{GF_with_S_site}
\end{eqnarray}
in which the replicated site action for site $i$ is of the form
\begin{equation}
S_{site}^{i}=S_{imp}+\sum_{\alpha w}\lambda_{w}\left(\bar{c}_{iw}^{\alpha}f_{iw}^{\alpha}+\bar{f}_{iw}^{\alpha}c_{iw}^{\alpha}\right).
\label{S_site}
\end{equation}
%
%
The inter-site coupling in Eq. $(\ref {GF_with_S_site})$ has been transferred to a coupling 
between dual fermions. This allows us to integrate out the real fermion degrees of freedom from the local site 
action $S_{site}^{i}$ for each site $i$ separately, i.e.,

\begin{eqnarray}
& &\int\prod_{\alpha w}d\bar{c}_{i}^{\alpha}dc_{i}^{\alpha}e^{-S_{site}[\bar{c}_{i}^{\alpha},c_{i}^{\alpha};\bar{f}_{i}^{\alpha},f_{i}^{\alpha}]} \nonumber \\
&=& Z_{imp} e^{-\sum_{w\alpha}\lambda_{w}^{2} g_{imp}^{\alpha}(w)\bar{f}_{iw}^{\alpha}f_{iw}^{\alpha}-\sum_{\alpha\beta}V_{d,i}^{\alpha,\beta}[\bar{f}_{i}^{\alpha},f_{i}^{\beta}]},
\label{Vdf_def}
\end{eqnarray}
in which $Z_{imp}$ is the partition function for the replicated impurity system.
As in the clean case ~\cite{Rubtsov08},  formally this can be done up to infinite order, which makes mapping to 
DF $exact$. Choosing for convenience $\lambda_{w}=g_{imp}^{\alpha}(w)^{-1}$, the lowest-order of the replicated 
non-antisymmetrized DF potential $V_{d,i}^{\alpha,\beta}[\bar{f}_{i}^{\alpha},f_{i}^{\beta}]$ reads as
\begin{equation}
V_{d,i}^{\alpha,\beta}[\bar{f}_{i}^{\alpha},f_{i}^{\beta}]= \frac {1}{2}\sum_{w w'}\gamma ^{\alpha,\beta}(w,w')\bar{f}_{iw}^{\alpha}\bar{f}_{iw'}^{\beta}f_{iw'}^{\beta}f_{iw}^{\alpha},
\end{equation}
where the CPA full vertex 
\begin{equation}
\gamma^{\alpha,\beta}(w,w')=\frac{-\chi_{imp}^{\alpha\beta}(w,w')-\chi_{0,imp}^{\alpha\beta}(w,w')}{[g_{imp}^{\alpha}(w) g_{imp}^{\beta}(w')]^{2}},
\end{equation}
\begin{equation}
\chi_{0,imp}^{\alpha\beta}(w,w') =  -g_{imp}^{\alpha}(w) g_{imp}^{\beta}(w^\prime),
\end{equation}
with $g_{imp}^{\alpha}(w) =-\displaystyle \int\mathcal{D}\bar{c}\mathcal{D}c\,e^{-S_{imp}}c_{w}^{\alpha}\bar{c}_{w}^{\alpha}$
and $\chi_{imp}^{\alpha\beta}(w,w')=\displaystyle \int \mathcal{D}\bar{c} \mathcal{D}c\,e^{-S_{imp}}c_{w}^{\alpha}c_{w'}^{\beta}\bar{c}_{w'}^{\beta}\bar{c}_{w}^{\alpha}$ being the impurity averaged single- and two-particle Green functions, respectively.

After taking the derivative with respect to the source field $\eta _{wk}$, the Green function of Eq.~$(\ref{GF_with_S_site})$ reads as
\begin{equation}
\left<G_{k}(w)\right>_{av}  = \left(\Delta_{w}-\varepsilon_{k}\right)^{-1}+\frac{\left<G_{d,k}(w)\right>_{av}}
{[\left(\Delta_{w}-\varepsilon_{k}\right)g_{imp}^{\alpha}(w)]^{2}},
\end{equation}
where we define the averaged DF Green function as 
\begin{eqnarray}
\left<G_{d,k}(w)\right>_{av}&=&-\lim_{m\rightarrow0}\frac{1}{m}\sum_{\alpha^\prime}\int \mathcal{D}\bar{f} \mathcal{D}f\, e^{-\sum_{wk\alpha}S_{d}^{0}}
\nonumber \\
&\times &e^{-\sum_{i\alpha\beta }V_{d,i}^{\alpha,\beta}[\bar{f}_{i}^{\alpha},f_{i}^{\beta}]}f_{wk}^{\alpha^\prime}\bar{f}_{wk}^{\alpha^\prime},
\label{GD}
\end{eqnarray}
and $S_{d}^{0}=\bar{f}_{wk}^{\alpha}
\left[\displaystyle \frac {(\Delta_{w}-\varepsilon_{k})^{-1}+g_{imp}^{\alpha}(w)}{g_{imp}^{\alpha}(w)^{2}}\right]
f_{wk}^{\alpha}$ is the non-interacting DF action.

\begin{figure}[h!]
\includegraphics[scale=0.6]{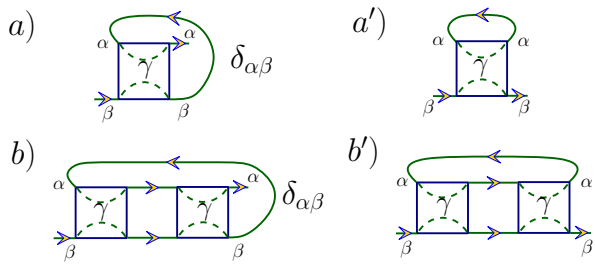}\caption{(color online) First and  second order DF self-energy 
diagrams. All diagrams with scattering of electrons along a closed loop, e.g. $a'$ and $b'$, vanish in the replica limit. 
Here, $\gamma$ is the CPA full vertex, with conserved frequency indicated by dashed lines and $\alpha$ and $\beta$ 
the replica.}
\label{Fig1: vertex}
\end{figure}

To calculate the DF Green function of Eq.~$(\ref {GD})$, we use diagrammatic perturbation theory. Here the non-trivial 
and crucial difference between the disordered and clean cases is that the interaction between replicas is off-diagonal, 
which puts certain constraints on the topology of Green function graphs. In particular, 
all graphs with closed fermion loops vanish (Fig. \ref {Fig1: vertex}). This is because 
each closed fermion loop contains one free replica summation which gives an extra factor of $m$ in Eq.~$(\ref {GD})$, 
and thus equals to zero when ${m\rightarrow 0}$ ~\cite{Atland}.

\paragraph*{Single-particle properties.-}
We first present the single-particle Green function for a one-dimensional lattice.  After solving the impurity part, 
one obtains the averaged impurity Green function $g_{imp}(w)=\displaystyle \int dVp(V)\frac{1}{iw_n +\mu-\Delta(w)-V}$ 
and corresponding impurity vertex $\gamma$.
Then, the DF part is solved self-consistently using standard diagrammatic perturbation theory. 
Next, the real lattice Green function from Eq.~$(\ref{GD})$ is recalculated, and new hybridization function $\Delta(w)$ is constructed 
to parametrize the impurity problem. This is repeated until self-consistency is reached, namely $\sum_k G_{d,k}(w)=0$, with all 
local diagrams (e.g. diagram a in Fig. \ref{Fig1: vertex}) being zero.

\begin{figure}[h!]
\includegraphics[width=3.3in]{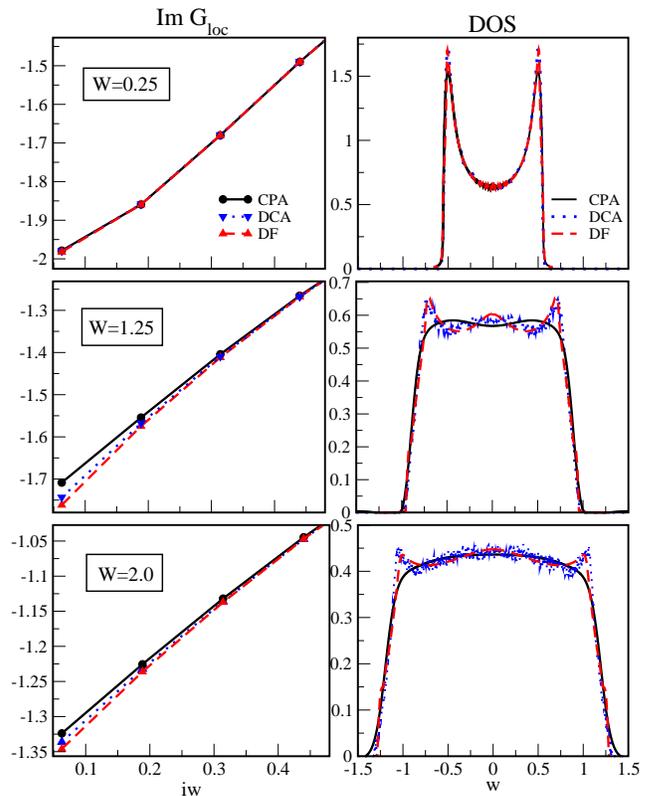}
\caption{(color online) The imaginary part of the local Matsubara Green function at $T=0.02$ (left) 
and the total density of states (right) for different disorder strengths: $W=0.25,1.25, 0.2$ ($4t=1$). 
For comparison, we present data obtained within CPA, a finite cluster DCA $(N_c=20)$ and DF methods. 
Inclusion of inter-site correlations leads to corrections to the CPA Green 
function (left panel) and appearance of additional structures at larger disorder in the total density 
of states (right panel).  In each case, the DF captures the features of the DCA density of states and 
is in nearly exact agreement for the Green function.}
\label{Fig2-GF-dos}
\end{figure}

In Fig.~\ref{Fig2-GF-dos} we present results for single-particle Green function obtained from
a fully self-consistent infinite ladder diagram summation (in both particle-hole (p-h) and 
particle-particle (p-p) channels) for the DF self-energy. To consider the effect of non-local 
correlations, we compare our DF results with CPA and DCA results for cluster size $N_c=20$. 
The local Matsubara Green function (left panel) shows that inclusion of inter-site correlations 
leads to corrections to the CPA Green function. Both DF and DCA results show good agreement at 
small and large disorder strength.  The local density of states (DOS) (right panel) also 
displays satisfactory agreement between DF and DCA results. Indeed, for weak disorder ($W=0.25$), 
the results from  CPA, DCA and DF calculations are practically the same. As the disorder strength 
increases, the non-local corrections become important (with finite momentum dependence of the 
self-energy) and the differences between the CPA and DF DOS are more pronounced. The DF 
successfully captures such correlations by producing additional features \cite{a_gonis_92}
which are also in good agreement with the fully converged DCA result, especially for large 
disorder strength ($W=2.0$). 

\paragraph*{Two-particle properties.-}
Although the CPA provides a good qualitative description of single-particle quantities, it fails 
to capture Anderson localization because the two-particle vertex does not depend on the transfer 
momentum between incoming and outgoing particles. Thus, the conductivity reduces to the bare 
p-h bubble so that vertex corrections  are ignored \cite{{Vollhardt_Wolfle}, {Velicky}}.

\begin{figure}[t!]
\includegraphics[width=3.4in]{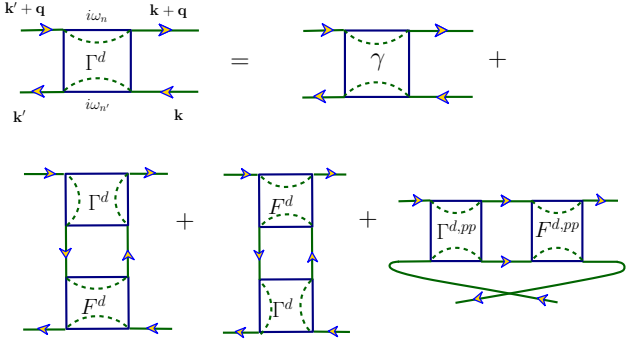}
\caption{(color online) Irreducible DF p-h horizontal vertex $\Gamma ^d$ is calculated using parquet 
equation with crossing contributions from p-h ``vertical'' and p-p channels. The fully 
irreducible vertex is approximated by $\gamma$. $\Gamma ^{d,pp}$ and $F^{d,pp}$ are the irreducible 
and full p-p vertices, respectively.} 
\label{Fig_parquet}
\end{figure}

In our scheme, the full vertex is non-local, so we expect to obtain finite vertex corrections  
describing ``weak'' localization effects \cite{LeeRMP}.  As our formalism is best converged on 
Matsubara frequency, we calculate the low temperature dc conductivity following \cite {Scalettar} as
\begin{equation}
\sigma_{dc}=\frac{\beta^{2}}{\pi}\Lambda_{xx}{ \left( {\bf q } =0,\tau=\frac{\beta}{2}\right)},\label{dc_cond}
\end{equation}
$\beta=1/k_BT$, and current-current correlation function 
$\Lambda_{xx}({\bf q},\tau)=\langle j_x({\bf q},\tau)j_x(-{\bf q},0)\rangle$. 
To obtain this lattice correlation function, one needs to calculate  the DF two-particle 
Green function $\chi^d=-\chi_0^d-\chi_0^d F^d\chi_0^d$, with $\chi_0^d=G^d G^d$~\cite{Rubtsov08}.  For the 
disordered case, one has to remember that in the DF vertex $F^d$ all diagrams containing closed loops are zero. 

As usual, $F^d$ is obtained from the Bethe-Salpeter equation $F^d=\Gamma^d +\Gamma^d \chi_0^d F^d$, where $\Gamma^d$ is 
the irreducible DF vertex in the p-h horizontal channel (c.f.\ Fig. \ref{Fig_parquet}). To calculate the last quantity, 
we use the parquet equations which account for the crossing contributions from the p-p and the ``vertical'' 
p-h channels \cite{Janis_parquet}. Here the fully irreducible vertex is approximated by the impurity 
full vertex $\gamma$. This procedure allows us to incorporate the important maximally-crossed diagrams \cite{Vollhardt_Wolfle}. 
The conductivity can be decomposed into two parts, $\sigma=\sigma_0+\Delta \sigma$, where $\sigma_0$ is 
the mean-field Drude conductivity, coming from the bare bubble $\chi_0$, and the second part $\Delta \sigma$ incorporates 
the vertex corrections. 

\begin{figure}[t!]
\includegraphics[width=3.3in]{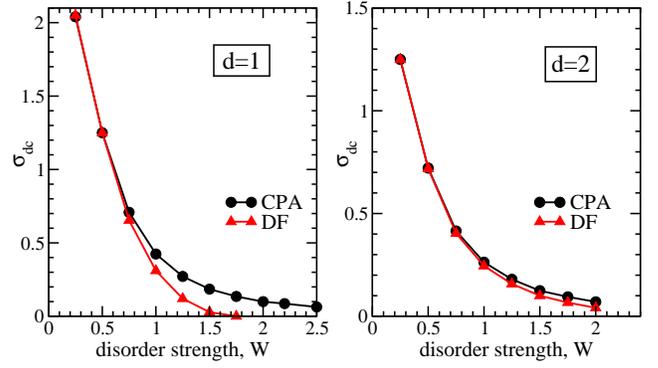}\caption{(color online) Conductivity as function of disorder 
strength using CPA and DF methods is shown. Data are obtained for $d=1$ and $d=2$ at $T=0.02$. 
Our results show that vertex corrections incorporated in DF approach  allow to 
capture weak localization leading to the decrease of the net conductivity.} 
\label{Fig_conductivity}
\end{figure}

Our results for the dc conductivity in dimensions $d=1$ and $d=2$ are presented in Fig. \ref{Fig_conductivity}. 
The data show that the disorder DF method is able to capture weak localization with vertex corrections 
(vanishing in CPA) leading to a net decrease of conductivity. In $d=1$, as the disorder strength increases, the DF 
vertex corrections are more pronounced, while in $d=2$ they are much weaker, as expected~\cite{LeeRMP}. 

\paragraph*{Conclusions.-}
We generalize the DF approach to include disordered non-interacting systems using the replica method. 
Our formalism incorporates non-local disorder-induced correlations neglected in the single-site CPA. 
While in the DCA the multi-scattering effects are limited by the size of the cluster, our method allows us to treat 
spatial correlations on all length scales. Comparing our results with large-cluster DCA data, we find rather good 
agreement. This shows that the DF scheme presents a powerful alternative to the existing cluster 
theories for description of non-local physics in disordered systems. More significantly, our method 
incorporates finite weak localization corrections to the mean-field conductivity - the precursor 
effect of Anderson localization. We believe that the DF disorder scheme traces a clear avenue to study a wide 
variety of physical phenomena, including the interplay of weak localization effects and strong electron 
interactions, which may be treated on equal footing in our method.  Work on this direction and generalization
to cluster cases~\cite{YangSXDF} are in progress. As a promising candidate to replace CPA, its application to study
 non-local effects in  electronic structure calculations is also envisioned.


\begin{acknowledgments}

We thank V. Janis and V. Dobrosavljevic for very useful discussions. This work is supported by the 
DOE BES CMCSN grant DE-AC02-98CH10886 (HT) and DOE SciDAC grant DE-FC02-06ER25792 (SY and MJ).  Additional support was 
provided by NSF EPSCoR Cooperative Agreement No. EPS-1003897 (ZM, JM). 
\end{acknowledgments}



\end{document}